\documentclass[dvipsnames]{IEEEtran}
\usepackage{cite}
\usepackage{amsmath,amssymb,amsfonts, xcolor, bm,here}
\usepackage{algorithmic}
\usepackage{soul}
\usepackage{graphicx}
\usepackage{textcomp}
\def\BibTeX{{\rm B\kern-.05em{\sc i\kern-.025em b}\kern-.08em
    T\kern-.1667em\lower.7ex\hbox{E}\kern-.125emX}}
   \usepackage{tikz}
\usetikzlibrary{arrows,shapes,calc}
\usepackage{pgfplots}
\usetikzlibrary{positioning}

\newtheorem{rem}{\sc Remark}
\begin{document}
\title{A Heuristic for Dynamic Output Predictive Control Design for Uncertain Nonlinear Systems}
\author{Mazen Alamir
\thanks{This work was supported by the MIAI @ GrenobleAlpes under Grant ANR-19-P31A-003.}
\thanks{The author is with the University of Grenoble Alpes, CNRS, Grenoble INP, GIPSA-lab, 38000 Grenoble, France (e-mail: mazen.alamir@grenoble-inp.fr). }}

\maketitle

\begin{abstract}
\color{black} In this paper, a simple heuristic is proposed for the design of uncertainty aware predictive controllers for nonlinear models involving uncertain parameters. The method relies on Machine Learning-based approximation of ideal deterministic MPC solutions with perfectly known parameters. An efficient construction of the learning data set from these off-line solutions is proposed in which each solution provides  many samples in the learning data. This enables a drastic reduction of the required number of Non Linear Programming problems to be solved off-line while explicitly exploiting the statistics of the parameters dispersion.  The learning data is then used to design a fast on-line output dynamic feedback that explicitly incorporate information of the statistics of the parameters dispersion. An example is provided to illustrate the efficiency and the relevance of the proposed framework. It is in particular shown that the proposed solution recovers up to 78\% of the expected advantage of having a perfect knowledge of the parameters compared to nominal design. 
\end{abstract}


\section{Introduction}\label{sec:introduction}
Over the last two decades, Nonlinear Model Predictive Control (NMPC) \cite{Mayne2000} spread in all domains and became a first choice when it comes to design control feedbacks for systems that admit faithful models. This is due to the maturity of its stability assessment as well as to the availability of user-friendly solvers \cite{Andersson2018}. As a direct consequence of this success, investigations started regarding the best way to extend the use of NMPC to systems that are represented through uncertain models. \\ \ \\ {\color{black} After some early attempts involving over stringent robust NMPC frameworks \cite{Magni2007}, robust adaptive MPC frameworks have been proposed  \cite{adetola2011, Guay2015} to provide on-line model adaptation for uncertain systems under the assumption of affine parameters, strict identifiability, persistent excitation and full state measurement.  More recently, the concept of Stochastic NMPC (SNMPC)  \cite{MAYNE2016184, Mesbah2018} emerged. In a nutshell, SNMPC frameworks implement the basic NMPC scheme with the standard deterministic cost and constraints being replaced by their expected values, considered as functions of the model's uncertainties. While the conceptual simplicity of this shift in paradigm is appealing, its consequence on the computational burden appears as a \emph{deal breaker} despite some recent attempts \cite{SCHILDBACH20143009, LUCIA20131306, Alamir2020SMPC}. Indeed, SNMPC requires the on-line computation of the expectation of nonlinear functions of several variables. This computation is to be performed for any value of the decision variable that is encountered during the on-line iterations of the optimization algorithm.  This complexity justified an increasing interest in off-line computation based solutions. \\ \ \\ One of these is related to Stochastic Dynamic Programming (SDP) / Reinforcement Learning (LR) alternatives \cite{Mesbah2018, BertsekasRL2019}  which offer elegant solutions for small-size problems. Another research track is related to the approximation of the {\em appropriate} optimal feedback/strategy via Deep/Machine learning-based fitting schemes (see \cite{LUCIA2018511, karg2019probabilistic} and the references therein). More precisely, a cloud of initial states is created and for each element in the cloud, a multi-stage SNMPC is solved (this needs finite and generally small number of possible uncertainty values to be allowed). Then a neural network is trained via deep-learning in order to fit a function that maps the initial state to the stochastically optimal feedback control and or strategy. This approach leads to extremely heavy off-line computational burden without providing theoretical guarantees, not only because of the inherent and unavoidable data incompleteness which is a shared issue by all data-driven solutions (including the one proposed in this paper), but also because all these schemes mainly use the multi-stage SNMPC framework which needs the uncertainty realizations to be limited to a few number of possible values which obviously scarcely fits the real-life situations. Consequently, these approaches should be viewed as {\em smart and reasonably tractable \underline{heuristics}} to address a problem that admits no rigorously computable solution}. \\ \ \\ 
It is not the intention of this paper to propose a {\em better framework} than the ones cited above and the references therein. This is because the comparison between {\em heuristics} is a complex multi-criteria based process that would go beyond the scope of this short contribution. In particular, while the scheme proposed hereafter is appealing from a computational point of view, its scope of application is limited to nonlinear models with fixed (or slowly varying) uncertain parameters as main source of uncertainty. In this particular but highly relevant case, the following intuition motivates this paper:\\ \ \\ 
\begin{minipage}{0.02\textwidth}
\color{black} \rule{1mm}{22mm} 
\end{minipage}
\begin{minipage}{0.44\textwidth} 
\color{black} under specific circumstances, the deterministic optimal solution that would have been computed, \underline{should the parameters be known}, can be learned from the previous measurement over some observation window.
\end{minipage}
\ \\ \ \\ \ \\
{\color{black} In the ideal case, this statement results from a perfect identifiability of the pair of state and parameter vectors leading to the optimal MPC control being an exact function of the previous measurement $\bm y^{(-)}_k$:
\begin{align}
\hat{\bm u}^\star\bigl(\bm y^{(-)}_k\bigr) = &\ \mbox{arg}\min_{\bm u\in \mathbb U^N} J\bigl(\bm u\ \vert\ \hat x(\bm y^{(-)}_k), \hat w(\bm y^{(-)}_k) \bigr) 
\end{align}
making $\bm y_k^{(-)}$ and $\hat u^\star_k$ ideal candidates to be viewed respectively as features vector and a label in a Machine Learning (ML) identification step. 
When extended observability does not hold, the above relationship can be used with the understanding that  $\hat x(\bm y^{(-)}_k)$ and $\hat w(\bm y^{(-)}_k)$  are {\bf non uniquely determined quantities} since for a given instance of the measurement profile $\bm y_k^{(-)}$, there is a cloud of possible values of $(\hat x, \hat w)$ and hence a cloud of possibly indistinguishable optimal control inputs $\hat u^\star$. The data generation using the statistics of the dispersion of $w$ and the following ML identification step helps making a statistically rational choice among the elements of this cloud.}\\ \ \\ 
Following the above lines, this paper proposes a framework for the design of uncertainty-aware output feedback law which is fitted from learning data that is constructed in a computationally efficient manner. The latter involves two main differences with standard approaches: 1) the off-line computation involves only deterministic optimal control problems. and can therefore directly benefit from the state-of-the-art available NMPC solvers.  2) Each single solution is used to extract multiple instances (that can be as high as hundreds) to be included in the learning data set. {\color{black} This is done by exploring the open-loop optimal optimal trajectories and concatenate the sub-optimal instances of $(\bm y^{(-)}_{k+i}, u^\star_{k+i}(\bm y_k^{(-)}))$ rather than collecting a single pair}. \\ \ \\ 
This paper is organized as follows: Section \ref{sec-def} presents some definitions and notation and states precisely the problem under study. The proposed framework is detailed in Section \ref{sec-proposed} together with the working assumptions that might induce its success. An illustrative example is proposed in Section \ref{sec-validation} in order to show the steps of the framework and evaluate the impact of parameters choice on the quality of the results. Finally, Section \ref{sec-conc} concludes the paper and gives some hints for further investigations. 
\section{Definition, notation and problem statement} \label{sec-def}
Let us consider nonlinear systems governed by the following discrete-time dynamics:
\begin{align}
x_{k+1}&=f(x_k,u_k,w) \label{syst1}\\
y_k&= h(x_k,u_k,w)\label{syst2}
\end{align}
where (\ref{syst1}) and (\ref{syst2}) describe the dynamics and the measurement equation respectively. The notation $x_k\in \mathbb{R}^{n_x}$, $u_k\in \mathbb U\subset \mathbb{R}^{n_u}$ and $y_k\in \mathbb{R}^{n_y}$ stand for the state, input and measured output vectors at instant $k$. The vector of parameters $w\in \mathbb{R}^{n_w}$ is supposed to be constant although uncertain. Moreover, it is assumed  that $w$ belongs to some bounded subset $\mathbb W$ over which a probability density function (pdf) $\mathcal W$ is known so that a statistically relevant samples of $w$ can be drawn if required. {\color{black} It is also assumed that a set of states $\mathbb X$ of interest is considered with some random sampling rule which reflects some form of relative importance or somehow relevance\footnote{In the absence of any knowledge, uniform distribution over some hyper-box can be simply used.}}. In what follows, a boldfaced notation refer to signals profiles over some past or future time windows, in particular the notation $\bm u:=(u_0^T, u_2^T, \dots, u_{N-1}^T)^T\in \mathbb U^N$ denotes a sequence of future control actions over some prediction horizon of length $N$ while $\bm y^{(-)}_k$  refers to the sequence of $M+1$ previous measurement vectors, namely
\begin{equation}
\bm y^{(-)}_k := \begin{bmatrix}
y_k\cr y_{k-1}\cr 
\vdots\cr 
y_{k-M}
\end{bmatrix} \in \bigl[\mathbb{R}^{n_y}\bigr]^{M+1} \label{defdeybar}
\end{equation}
The notation $$\bm x^{(\bm u)}(x_0, w)= \begin{bmatrix}
x_0^{(\bm u)}(x_0, w)\cr 
x_1^{(\bm u)}(x_0, w)\cr 
\vdots \\
x_N^{(\bm u)}(x_0, w)
\end{bmatrix}\in \Bigl[\mathbb{R}^{n_x}\Bigr]^{N+1} $$ denotes the state trajectory under (\ref{syst1}) when the control profile $\bm u$ and when the parameter vector $w$ is used, namely:
\begin{align*}
x_0^{(\bm u)}(x_0,w)&=x_0\\ 
x_{k+1}^{(\bm u)}(x_0,w)&=f(x_k^{(\bm u)}(x_0,w), u_k, w)
\end{align*}
It is assumed that the control objective would be defined  through a cost function of the form 
\begin{equation}
J(\bm u\ \vert\ x_k,w):= \sum_{i=1}^{N} \ell_i\bigl(x_{i}^{(\bm u)}(x_k), u_{i-1}\bigr) \label{cost}
\end{equation}
should the current state $x_k$ and the parameter vector $w$ be known. The design problem that is addressed in the present paper can be shortly stated as follows:
\begin{center}
\tikz{
\node[rounded corners, fill=Gray!10, inner xsep=5mm, inner ysep=5mm]at(0,0)(A){
\begin{minipage}{0.4\textwidth}
Design a dynamic output feedback of the form:
$$u_k:=K(\bm y^{(-)}_k)$$
that associates to each past measurement profile $\bm y^{(-)}_k$ an associated input in such a way that incorporates the presumably known statistics $\mathcal W$ and that is oriented towards the optimization of the cost function $J$. 
\end{minipage}
};
\node[above] at(A.north){\sc Uncertainty-aware dynamic output  feedback};
}\label{obsprob}
\end{center}
\begin{rem}
\em 
it is assumed that all the constraints except the control saturation related ones (which are expressed through $\mathbb U$) are included in the very definition of the stage cost map $\ell$ for the sake of simplicity of exposition. Moreover, it should be underlined that the specific structure of the cost function, being the sum of decoupled stage-cost terms is not explicitly needed although it is kept because it fits all the available optimal control problems' solvers that we rely on in the design step. 
\end{rem}
\section{The proposed framework} \label{sec-proposed}

\subsection{The design viewed as a map identification problem}
The starting point lies in the fact that {\color{black} the problem reduces to a standard deterministic problem} if the pair $(x_k,w)$ of state and parameter vector were reconstructible via {\bf  extended observation} \cite{besancon2007nonlinear}, namely, if one can reconstruct estimations $\hat x_k$ and $\hat w$ of these two quantities as functions of the previous measurement:
\begin{equation}
\hat x(\bm y^{(-)}_k)\quad;\quad \hat w(\bm y^{(-)}_k)
\end{equation}
Indeed, in this case, the answer to the problem stated in the previous section would obviously be given by:
\begin{equation}
K(\bm y^{(-)}_k):=  \hat u^\star_0\bigl(\bm y^{(-)}_k\bigr) \label{ijh67}
\end{equation}
where $\hat u^\star_0(\bm y^{(-)}_k)$ is the first control in the optimal sequence that solves the following optimization problem:
\begin{align}
\hat{\bm u}^\star\bigl(\bm y^{(-)}_k\bigr) = &\ \mbox{arg}\min_{\bm u\in \mathbb U^N} J\bigl(\bm u\ \vert\ \hat x(\bm y^{(-)}_k), \hat w(\bm y^{(-)}_k) \bigr) 
\end{align}
{\color{black} When extended observability does not hold, the pairs  $\hat x(\bm y^{(-)}_k), \hat w(\bm y^{(-)}_k)$  are {\bf non uniquely determined quantities} since for a given instance of the measurement profile $\bm y_k^{(-)}$, there is a cloud of possible values of $(\hat x, \hat w)$ and hence a cloud of possibly indistinguishable optimal control inputs $\hat u^\star$. The data generation using the statistics of the dispersion of $w$ and the following ML identification step helps making a statistically rational choice among the elements of this cloud.}\\ \ \\ 
 To summarize:
\begin{center}
\tikz{
\node[rounded corners, fill=Gray!10, inner xsep=5mm, inner ysep=5mm]at(0,0)(A){
\begin{minipage}{0.4\textwidth}
{\color{black}  Given the statistical dispersion of the parameters, fit the best map $F(\bm y^{(-)}_k)\approx \hat u^\star_0(\cdot)$ that links the sequence of previous measurement $\bm y^{(-)}_k$ to the optimal control $u_k=F(\bm y^{(-)}_k)$. }
\end{minipage}
};
\node[above] at(A.north){\sc The map to be identified};
}\label{obsprob}
\end{center}
Now the question is: \\ \ \\ 
\begin{minipage}{0.03\textwidth}
\ 
\end{minipage}
\begin{minipage}{0.01\textwidth}
\color{black} \rule{1mm}{13mm} 
\end{minipage}
\begin{minipage}{0.44\textwidth}
{\em  \color{black} How to build a relevant and rich learning data that can be used to fit the map $F$ while reducing as far as possible the amount of off-line computation ?}
\end{minipage}
\subsection{Building the learning data}
\begin{figure}[H]
\begin{center}
\begin{tikzpicture}[scale=0.8]
\draw[<->] (0,3) -- (0,0) -- (9,0) node[below]{\footnotesize time};
\draw[fill=black] (0,1.2) circle (2pt) node[left]{\footnotesize $x_0$}; 
\foreach \x in {0,...,19}
\draw[dotted] (\x/20*9, 0) -- (\x/20*9, 3);
\node[below] at(0,0){\footnotesize $0$};
\node[below] at(8,0){\footnotesize $N$};
\draw[thick] (8.1,0) -- (8.1,3);
\node[below] at(2.25,0){\footnotesize $M$};
\draw[thick] (2.25,0) -- (2.25,3);
\draw[<->] (0,2) -- node[midway, below] {\footnotesize $[\bm y^\star(q)]^{(-)}_M$}(2.25, 2);
\draw[<->] (0,2) -- node[midway, above] {\footnotesize $\bm y^{(-)}=$}(2.25, 2);
\draw[<->, Blue] (0,0.5) -- node[above] {\footnotesize\hskip 18mm Optimal trajectories $\bm u^\star(q)$}(8.1, 0.5);
\draw[Red, very thick] (2.25,1.8) -- (2.7, 1.8);
\draw[Black, very thick] (0,1) -- (0.45, 1);
\draw[Black, very thick] (0.45,1.2) -- (0.9, 1.2);
\draw[Black, very thick] (0.9,0.6) -- (1.35, 0.6);
\draw[Black, very thick] (1.35,0.8) -- (1.8, 0.8);
\draw[Black, very thick] (1.8,1.1) -- (2.25, 1.1);
\draw[Black, very thick] (2.7,1.7) -- (3.15, 1.7);
\draw[Black, very thick] (3.15,1.5) -- (3.6, 1.5);
\draw[Black, very thick] (3.6,1.45) -- (4.05, 1.45);
\draw[Black, very thick] (4.05,1.4) -- (4.5, 1.4);
\draw[Black, very thick] (4.5,1.3) -- (4.95, 1.3);
\draw[Black, very thick] (4.95,1.3) -- (8.1, 1.3);
\draw[Blue] (2.4, 1.8) edge [bend left]  ++(3,1.5);
\node[right, Blue] at(5.4, 3.3){\footnotesize $\hat u^\star_0(\bm y^{(-)}):=u^\star_M(q)$};
\end{tikzpicture}
\end{center}
\caption{\color{black} Generation of a sample pair $(\bm y^{(-)}, \hat u^\star_0(\bm y^{(-)})$ by solving an optimal control problem for a given sampled pair $q=(x_0,w)$. } \label{lasituationMIH}
\end{figure}
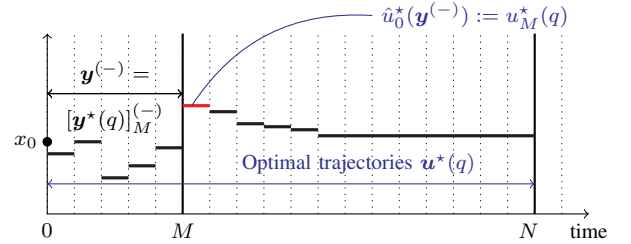

\begin{figure}
\begin{center}
\begin{tikzpicture}[scale=0.8]
\draw[<->] (0,3) -- (0,0) -- (9,0) node[below]{\footnotesize time};
\draw[fill=black] (0,1.2) circle (2pt) node[left]{\footnotesize $x_0$}; 
\foreach \x in {0,...,19}
\draw[dotted] (\x/20*9, 0) -- (\x/20*9, 3);
\node[below] at(0,0){\footnotesize $0$};
\node[below] at(8,0){\footnotesize $N$};
\draw[thick] (8.1,0) -- (8.1,3);
\node[below] at(2.25,0){\footnotesize $M$};
\draw[thick] (2.25,0) -- (2.25,3);
\draw[<->] (0,2) -- node[midway, above] {\footnotesize $\bm y^{\star (-)}_M$}(2.25, 2);
\draw[<->, Blue] (0,0.5) -- node[above] {\footnotesize\hskip 18mm Optimal trajectories $\bm u^\star(q)$}(8.1, 0.5);
\draw[Red, very thick] (2.25,1.8) -- (2.7, 1.8);
\draw[Black, very thick] (0,1) -- (0.45, 1);
\draw[Black, very thick] (0.45,1.2) -- (0.9, 1.2);
\draw[Black, very thick] (0.9,0.6) -- (1.35, 0.6);
\draw[Black, very thick] (1.35,0.8) -- (1.8, 0.8);
\draw[Black, very thick] (1.8,1.1) -- (2.25, 1.1);
\draw[Black, very thick] (2.7,1.7) -- (3.15, 1.7);
\draw[Black, very thick] (3.15,1.5) -- (3.6, 1.5);
\draw[Black, very thick] (3.6,1.45) -- (4.05, 1.45);
\draw[Black, very thick] (4.05,1.4) -- (4.5, 1.4);
\draw[Black, very thick] (4.5,1.3) -- (4.95, 1.3);
\draw[Black, very thick] (4.95,1.3) -- (8.1, 1.3);
\draw[Blue] (2.4, 1.8) edge [bend left]  ++(3,1.5);
\node[right, Blue] at(5.4, 3.3){\footnotesize $\hat u^\star_0(\bm y^{\star (-)}_M):=u^\star_M(q)$};
\end{tikzpicture}
\end{center}

\begin{center}
\begin{tikzpicture}[scale=0.8]
\draw[<->] (0,3) -- (0,0) -- (9,0) node[below]{\footnotesize time};
\draw[fill=black] (0,1.2) circle (2pt) node[left]{\footnotesize $x_0$}; 
\foreach \x in {0,...,19}
\draw[dotted] (\x/20*9, 0) -- (\x/20*9, 3);
\node[below] at(0,0){\footnotesize $0$};
\node[below] at(8,0){\footnotesize $N$};
\draw[thick] (8.1,0) -- (8.1,3);
\node[below] at(2.7,0){\footnotesize $M+1$};
\draw[Black, very thick] (2.25,1.8) -- (2.7, 1.8);
\draw[thick] (2.7,0) -- (2.7,3);
\draw[<->] (0.45,2) -- node[midway, above] {\footnotesize $\bm y^{\star (-)}_{M+1}$}(2.7, 2);
\draw[<->, Blue] (0,0.5) -- node[above] {\footnotesize\hskip 18mm Optimal trajectories $\bm u^\star(q)$}(8.1, 0.5);
\draw[Black, very thick] (0,1) -- (0.45, 1);
\draw[Black, very thick] (0.45,1.2) -- (0.9, 1.2);
\draw[Black, very thick] (0.9,0.6) -- (1.35, 0.6);
\draw[Black, very thick] (1.35,0.8) -- (1.8, 0.8);
\draw[Black, very thick] (1.8,1.1) -- (2.25, 1.1);
\draw[Red, very thick] (2.7,1.7) -- (3.15, 1.7);
\draw[Black, very thick] (3.15,1.5) -- (3.6, 1.5);
\draw[Black, very thick] (3.6,1.45) -- (4.05, 1.45);
\draw[Black, very thick] (4.05,1.4) -- (4.5, 1.4);
\draw[Black, very thick] (4.5,1.3) -- (4.95, 1.3);
\draw[Black, very thick] (4.95,1.3) -- (8.1, 1.3);
\draw[Blue] (2.85, 1.7) edge [bend left]  ++(2.55,1.5);
\node[right, Blue] at(5.4, 3.3){\footnotesize $\hat u^\star_0(\bm y^{\star (-)}_{M+1}):=u^\star_{M+1}(q)$};
\end{tikzpicture}
\end{center}

\begin{center}
\begin{tikzpicture}[scale=0.8]
\draw[<->] (0,3) -- (0,0) -- (9,0) node[below]{\footnotesize time};
\draw[fill=black] (0,1.2) circle (2pt) node[left]{\footnotesize $x_0$}; 
\foreach \x in {0,...,19}
\draw[dotted] (\x/20*9, 0) -- (\x/20*9, 3);
\node[below] at(0,0){\footnotesize $0$};
\node[below] at(8,0){\footnotesize $N$};
\draw[thick] (8.1,0) -- (8.1,3);
\node[below] at(3.15,0){\footnotesize $M+2$};
\draw[thick] (3.15,0) -- (3.15,3);
\draw[<->] (0.9,2) -- node[midway, above] {\footnotesize $\bm y^{\star (-)}_{M+2}$}(3.15, 2);
\draw[<->, Blue] (0,0.5) -- node[above] {\footnotesize\hskip 18mm Optimal trajectories $\bm u^\star(q)$}(8.1, 0.5);
\draw[Black, very thick] (0,1) -- (0.45, 1);
\draw[Black, very thick] (0.45,1.2) -- (0.9, 1.2);
\draw[Black, very thick] (0.9,0.6) -- (1.35, 0.6);
\draw[Black, very thick] (1.35,0.8) -- (1.8, 0.8);
\draw[Black, very thick] (1.8,1.1) -- (2.25, 1.1);for
\draw[Black, very thick] (2.7,1.7) -- (3.15, 1.7);
\draw[Red, very thick] (3.15,1.5) -- (3.6, 1.5);
\draw[Black, very thick] (3.6,1.45) -- (4.05, 1.45);
\draw[Black, very thick] (4.05,1.4) -- (4.5, 1.4);
\draw[Black, very thick] (4.5,1.3) -- (4.95, 1.3);
\draw[Black, very thick] (4.95,1.3) -- (8.1, 1.3);
\draw[Blue] (3.3, 1.5) edge [bend left]  ++(2.15,1.5);
\node[right, Blue] at(5.4, 3.3){\footnotesize $\hat u^\star_0(\bm y^{\star(-)}_{M+2}):=u^\star_{M+2}(q)$};
\end{tikzpicture}
\end{center}	
\caption{\color{black} Construction of the learning data $\mathcal D(q)$: by going forward in a moving window along the optimal trajectory computed for a single pair $q=(x_0,w)$ it is possible to generate a high number of different samples of pair $(\bm y^{(-)}, u)$ that can be used in the construction of the learning data.} \label{recedingHsampling}
\end{figure}
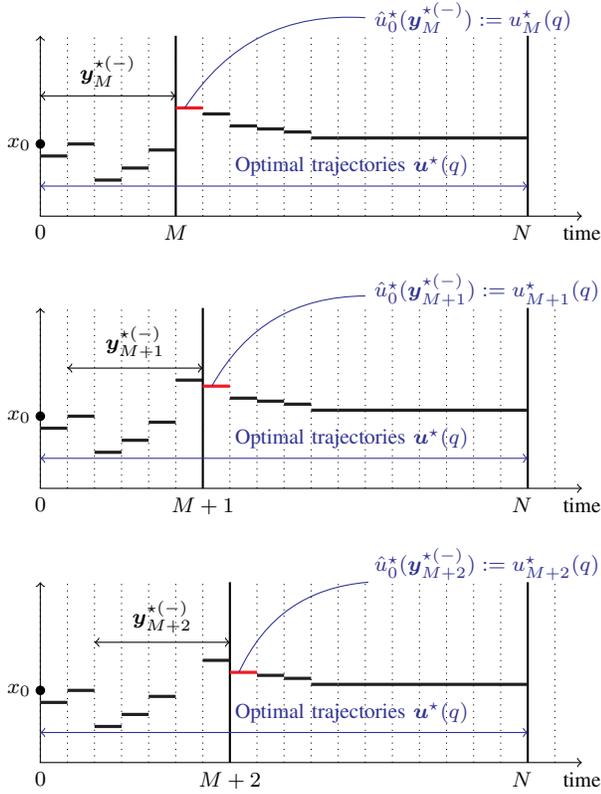

{\color{black} Consider the following procedure that is the elementary bloc in the data generation step}:
\begin{enumerate}
\item  Choose a long prediction horizon $N$.
\item Choose an observation horizon $M<N$.
\item Randomly choose a pair $q=(x_0,w){\color{black} \in \mathbb{R}^{n_x}\times \mathbb{R}^{n_w}}$
\item Use an efficient NLP solver (for instance \texttt{IPOPT-Casadi} \cite{Andersson2018}) to compute a minimizer $\bm u^\star(q)$ to $J(\cdot\ \vert\  q)$. One gets the situation depicted in Figure \ref{lasituationMIH}. 
\item {\color{black} This item explains how $m$ samples in the learning data is created using the optimal open-loop trajectories computed in the previous item for the specific value of $q=(x_0,w)$}. Indeed, consider the pair (see Figure \ref{lasituationMIH}):
\begin{equation}
\Bigl(\bm y^{(-)}, \hat u^\star_0(\bm y^{(-)})\Bigr):=\Bigl([\bm y^\star(q)]^{(-)}_M\ ,\ u^\star_M(q)\Bigr) \label{ghgFR95}
\end{equation}
\end{enumerate}
{\color{black} The key idea is that the pair defined in (\ref{ghgFR95}) approximates a pair of the form (\ref{ijh67}) provided that the prediction horizon $N$ is sufficiently long}. Indeed, $\bm y^{(-)}:=[\bm y^\star(q)]^{(-)}_M$, considered at instant $M$ is the vector of past measurements. It therefore contains the information regarding the pair $(x^\star_M(x_0),w)$ of current state and parameter vector. 
Now it is true that in order for (\ref{ghgFR95}) to be of the form (\ref{ijh67}), the following approximation:
\begin{equation}
u^\star_M(q)\approx u^\star_0(x^\star_M(q)) \label{Bellman}
\end{equation}
should be satisfied. But this approximation would have been a strict equality should $N$ be infinite thanks to the {\bf  Bellman principle}! Indeed, $u^\star_M(q)$ is the beginning of the {\em remaining part} of an optimal solution while $u^\star_0(x^\star_M(q))$ is the optimal solution of the updated problem and these two solutions would be identical should $N$ be infinite. \\ \ \\ 
Now the same argument holds if we iterate the process (still for the same $q$) using moving windows that end at instants $M+j$, for $j=1,\dots,m-1$. This means that by performing a single deterministic NLP solution, we can generate $m$ samples in the learning data, namely:
\begin{equation}
\mathcal D(q):= \Biggl\{\Bigl([\bm y^\star(q)]^{(-)}_k\ ,\ \bm u^\star_k(q)\Bigr)\Biggr\}_{k=M}^{M+m-1} \label{LD1}
\end{equation}
This procedure is shown in Figure \ref{recedingHsampling}.\\ \ \\ 
Repeating this procedure for a cloud of $n_q$ sampled values of the pair $q$, one gets a learning data:
\begin{equation}
\mathcal D := \Bigl\{\mathcal D(q^{(\ell)})\Bigr\}_{\ell=1}^{n_q} \label{LD2}
\end{equation}
containing $(n_q\cdot m)$ samples at the price of solving $n_q$ deterministic optimal control problems. 
\subsection{Performance evaluation}
The learning data  $\mathcal D$ can now be used to fit a model of the form\footnote{ML stands for (Machine Learning)-based model }:
\begin{equation}
u_k = \mbox{ML}(\bm y^{(-)}_k) \label{eqML}
\end{equation}
This feedback can be applied in closed-loop to a {\bf  new cloud} of values of $\bigl\{q^{(j)}=(x_0^{(j)},w^{(j)})\bigr\}_{j=1}^{n_v}$. Namely for each of these values, the feedback is implemented without knowledge of the associated $w^{(j)}$ (only the vector of past measurement is used through (\ref{eqML})). If the closed-loop is simulated during $N$ samples (equal to the prediction horizon), then the corresponding closed-loop cost, say $J_j^{cl}$ can be compared to the exact optimal cost $J^\star_j$ that can be viewed as the ideal solution (since deterministic problems are solved assuming that the parameter values $w^{(j)}$ is known). Obviously, if the optimization is perfectly done, one should systematically have $J^{cl}_j\le J^\star_j$ since any closed-loop sequence over the prediction horizon $N$ is naturally a candidate sequence of the open-loop optimization problem. 

\section{Numerical investigation} \label{sec-validation}
Consider the parallel reactor system \cite{Muller2014} commonly used in the study of deterministic Economic NMPC:
\begin{align}
\dot x_1&=1-w_1x_1^2e^{-1/x_3}-w_2x_1e^{-w_3/x_3}-x_1 \label{obsreac1D}\\
\dot x_2&=w_1x_1^2e^{-1/x_3}-x_2 \label{obsreac2D}\\
\dot x_3&=u-x_3 \label{obsreac3D}
\end{align}
where $x_1$ and $x_2$ stand for the concentrations of reactant and product respectively while $x_3$ represents the temperature of the mixture in the reactor. The control variable is given by the heat flow $u\in \mathbb U:=[0.049, 0.449]$. The objective of the control is to maximize the amount of product $x_2$ which is precisely the only measured variable. This leads to the following cost function:
\begin{equation*}
J(\bm u\ \vert\  x_0,w)=-\sum_{i=1}^{N}Cx_i^{(\bm u)}(x_0, w)\quad \mbox{\rm with}\quad C=(0,1,0)
\end{equation*}
In the forthcoming computations, a sampling period of $0.02$ time units is used. Based on this value and observing typical optimal trajectories, the prediction horizon length $N=250$ has been considered as it leads to a prediction horizon that is slightly greater than the settling time. The past measurement horizon length used to define $\bm y^{(-)}_k$ is taken equal to $M=10$.
\subsection{Generating the learning and validation data}
\begin{figure}
\begin{center}
\includegraphics[width=0.38\textwidth]{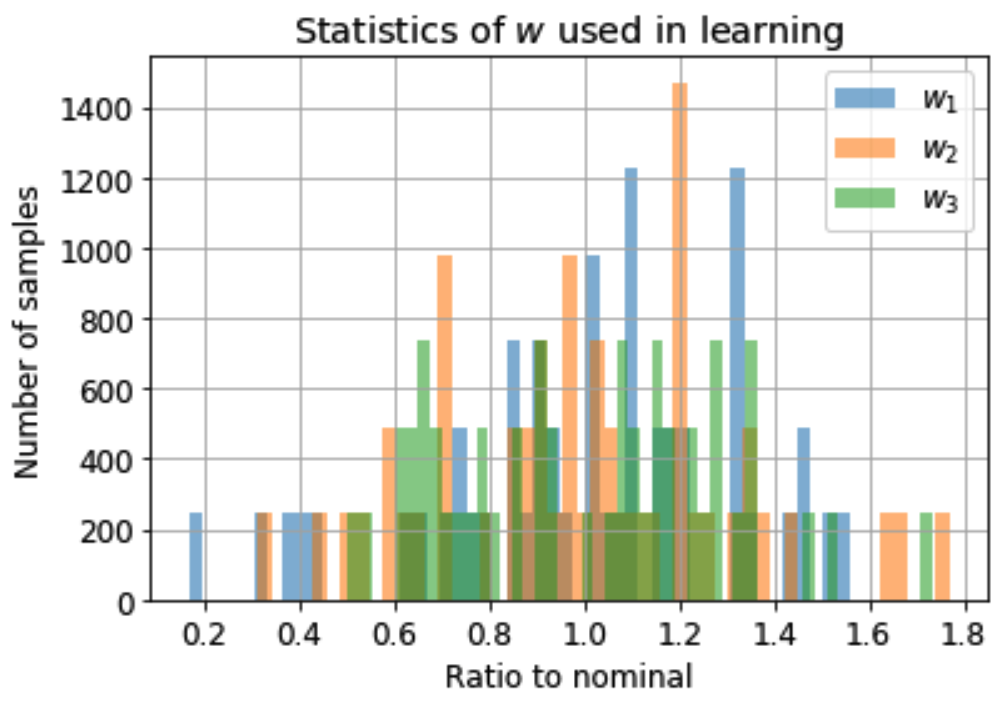}
\end{center}	
\caption{Dispersion of the parameters in the learning data in terms of ratio to the nominal components. (Ratios to nominal values are shown)}\label{statisticsw}
\end{figure}

\begin{figure}
\begin{center}
\includegraphics[width=0.4\textwidth]{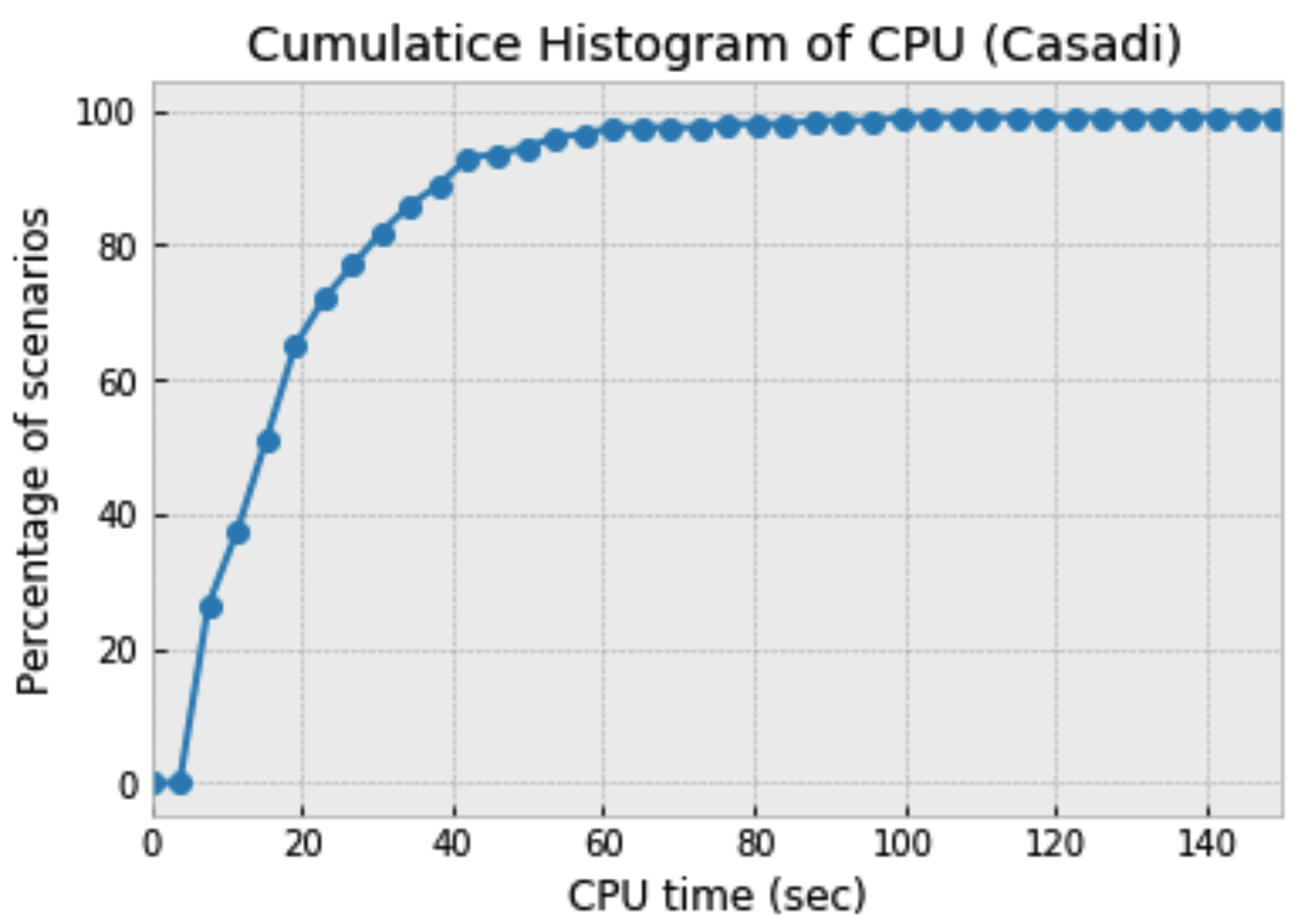}	
\end{center}
\caption{Cumulative histogram of computation time (sec) needed by IPOPT ({\bf multiple-shooting}) to solve a single optimal control problem when creating the learning data for the illustrative example.} \label{cpucasadi}
\end{figure}

\begin{figure}
\begin{center}
 \includegraphics[width=0.3\textwidth]{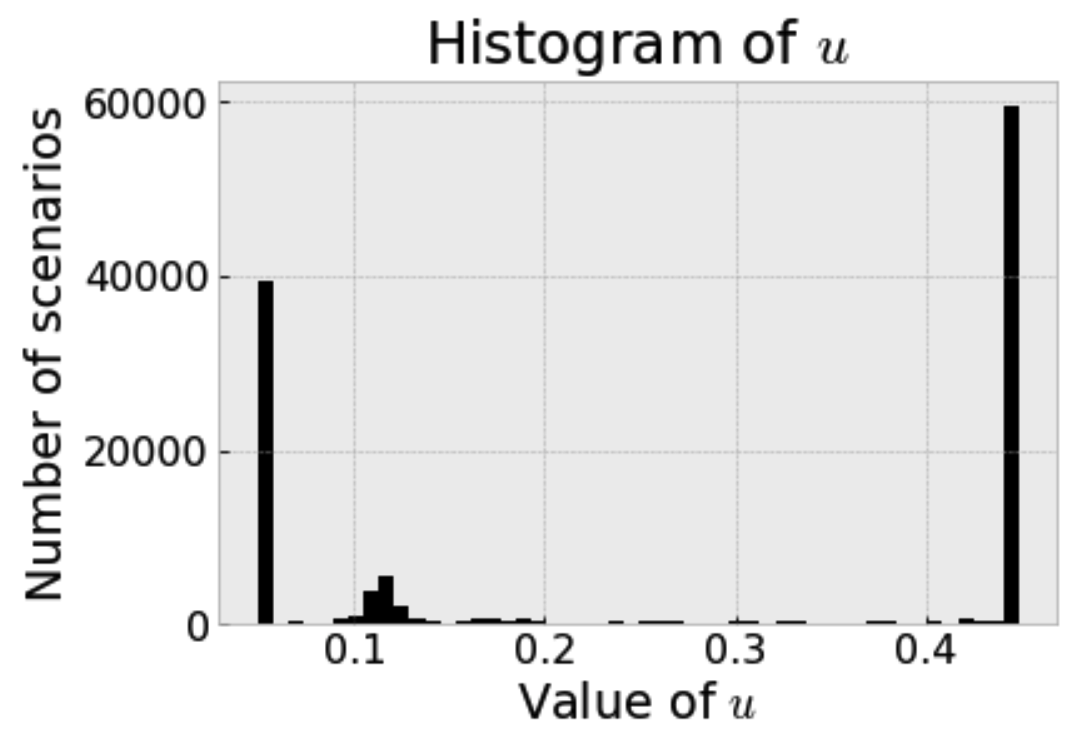}
\end{center}
\caption{Histogram of the values of the control $u$ present in the $500$ scenarios (of length $N=250$ each) contained in the learning data.} \label{histuoptdes}
\end{figure}

The initial states $x_0$ are uniformly sampled inside the set
\begin{equation}
\color{black} \mathbb{X}_0:= [10^{-4}, 0.5]\times [10^{-4}, 0.2]\times [10^{-4}, 0.25]
\end{equation}
which is known to include almost all relevant evolutions of the system. As for the values of the unknown parameter vector $w$, a normal distribution around the nominal value given by:
$$w^0=(10^4,400,0.55)$$
is used so that the random sampling is based on the rule:
\begin{equation}
w_i = (1+\nu_i)w^0_i\qquad \mbox{\rm with}\qquad \nu_i\in \mathcal N(0,\sigma_i) \label{wnormal}
\end{equation}
meaning that the true values are normally distributed around the nominal value with a known standard deviations. The values $\sigma_i=0.33$ are used for $i\in \{1,2,3\}$.
Figure \ref{statisticsw} shows the dispersion of the parameter values that results from the random sampling law (\ref{wnormal}). Note that the variations of the parameters span the interval going from 20\% to 180\% of the nominal values.
\  \\ \ \\
The learning data set is generated according to (\ref{LD1})-(\ref{LD2}) in which $n_q=500$ samples of pairs $(x_0,w)$ are drawn inside the above described sets. This leads to a learning set of cardinality $(500\cdot m)$ where $m$ is the number of times the moving window is translated in order to generate samples from a single solution a deterministic optimal control problem.  In what follows, several identification settings are tested for different values of $m\in \{10, 50, 100, 150, 200\}$ leading to learning data sets of different cardinality $\{5000, 25000, 50000, 75000, 100000\}$ samples having all in common almost the same computation time (the time needed to solve $n_q=500$ deterministic NLP problems). \\ \ \\ 
The cumulative histogram of computation time (in sec) needed by a multiple-shooting implementation of the optimal control solver \texttt{IPOPT} implemented using  \texttt{Casadi} is shown in Figure \ref{cpucasadi}.
The computation time shown in Figure \ref{cpucasadi} might seem too long for those who are commonly using \texttt{IPOPT} to solve standard regulation problems. Indeed, the median time is close de 15 sec! This is probably due to the oscillating character of optimal trajectory that is very specific to the underlying economic NMPC formulation and to the relatively long prediction horizon $N=250$ being used.  \\ \ \\ 
The histogram of control values over the prediction horizon of the $n_q=500$ sampled scenarios (including hence 125000 values) is shown in Figure \ref{histuoptdes}. This histogram suggests that classification tools are best suited for the learning of the control using ML tools compared to regression tools\footnote{for this specific example. }. Indeed, the following three-valued set of label values can be used:
{\color{black} 
\begin{equation}
\mbox{\rm label} = \left\{ 
\begin{array}{ll}
 1& \mbox{\rm if $u\le 0.075$}\\
 2& \mbox{if $u\in ]0.07, 0.14]$}\\
3& \mbox{otherwise}
\end{array}
\right. \label{lelabelyoptides}
\end{equation}}
with this definition of the label, it is possible to identify a {\bf  classifier} that associates to any time series of measurement $\bm y^{(-)}$ an element of the set $\{1,2,3\}$ and hence its corresponding control value in $\{0.049, 0.11, 0.449\}$. \\ \ \\ 
Note that if the distribution of the control values was more {\em continuously distributed} between $u^{min}=0.049$ and $u^{max}=0.449$, a regressor would have been more appropriate to address the identification problem. 
\subsection{Identifying the output feedback model}
\noindent {\color{black} The features vector is build using the previous profile of $x_2$ as output}. The Identification of the map ML$(\bm y^{(-)})$ has been obtained using a \texttt{RandomForestClassifier} from the \texttt{scikit-learn} freely available python library \cite{scikitlearn}. In order to avoid overfitting and enhance the quality of the extrapolation on unseen data, the \texttt{max\_leaf\_nodes} parameter has been set to $500$. 
\begin{figure}[H]
\begin{center}
 \includegraphics[width=0.23\textwidth]{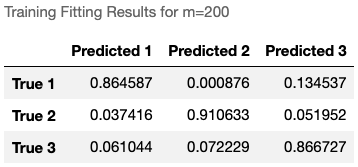}  \includegraphics[width=0.23\textwidth]{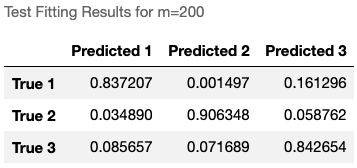} \\
 \includegraphics[width=0.23\textwidth]{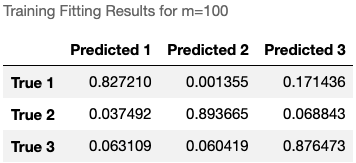}  \includegraphics[width=0.23\textwidth]{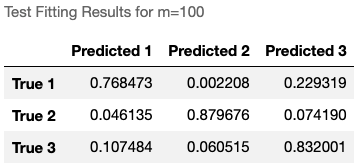} \\
 \includegraphics[width=0.23\textwidth]{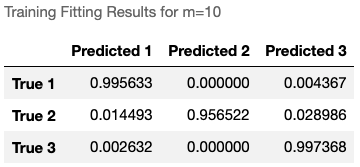}  \includegraphics[width=0.23\textwidth]{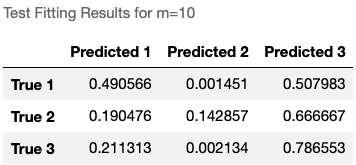} \\
\end{center}
\caption{Fitting results on the training and test data (split ratio (33\%) for the different values of receding horizon moves $m$ used in the construction of the learning data. Recall that the label the classifier tries to guess lies in $\{1,2,3\}$ as explained in (\ref{lelabelyoptides}).}\label{fittingresultsdifm}
 \end{figure}
The confusion matrices corresponding to the training and the test data (using a test-size ratio of 33\% of the learning data) are shown in Figure \ref{fittingresultsdifm} for $m\in \{10, 100, 200\}$. These matrices clearly show that when $m=10$ is used (cardinality of the learning set = 5000) the learning data is still not representative enough (as the precision on training data is significantly better than on test data). This clearly highlights by itself the relevance of the proposed heuristic since the computation time reported in Figure \ref{cpucasadi} suggests that the time that would be needed to generate $5000$ rigorously optimal although still insufficient data would be around $21$ hours while the time needed here to generate $100000$ samples is around 2h. 
\subsection{Performance evaluation}
\begin{figure}[H]
\begin{center}
\framebox{\includegraphics[width=0.4\textwidth]{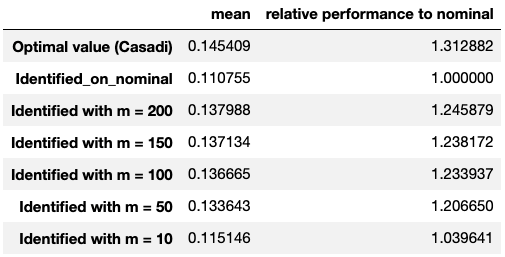}}
\end{center}
\caption{Comparison of the closed-loop performances on the validation data set. } \label{performanceDashOptimalChiffres}
\end{figure}

A new set of $100$ pairs of $(x_0,w)$ is generated and the $100$ initial states are used to generate learning data using the nominal value $w^0$ of the parameter vector. This is done in order to evaluate the benefit from explicitly learn the feedback from sampled parameter vector compared to a nominal design that only use the nominal expected value. \\ \ \\ Then for each value of $m$,  the corresponding identified feedback law is used to generate $100$ closed-loop  in which the newly drawn samples of the parameter vector (which are unknown to the previously identified feedbacks) is used in the simulated model. \\ \ \\ 
The results are shown in Figure \ref{performanceDashOptimalChiffres} where one can find 1) the ideal optimal closed-loop cost {\color{black} given by the NLP solver \texttt{IPOPT}} in the unrealistic case where the disturbance is perfectly known 2) the closed-loop performance obtained using the nominal feedback described above and 3) the closed-loop performance of the identified feedback using the different values of $m$. 
From Figure \ref{performanceDashOptimalChiffres}, the following observations can be made:
\begin{itemize}
\item  The optimal achievable values of the cost {\bf  when the parameter vector is perfectly known} is $31\%$ better (in average) than what the nominal identified controller enables to achieve. 
\item The proposed methodology enables to \hl{\bf  recover 78\% of the advantages\footnote{This is the ratio 0.245/0.312} of knowing perfectly the value of the parameter vector} when $m=200$ is used in the construction of the learning data. This corresponds to $24\%$ better average than the nominally optimal identified controller.  Almost similar results are obtained when $m\in \{150, 100\}$ is used. 
\item The average performance drops to $20\%$ ($66\%$ of recovered advantage) when $m=50$ is used while a drastic decrease corresponding to 3.9\% (and only $12\%$ of recovered advantage) when short $m=10$ is used in the building of the learning data. \hl{It is important to underline that if strict optimality is used in the data set building step (only one sample per NLP solution), the time that would be needed to solve the $100000$ NLP problems \footnote{Cardinality of the learning set for $m=200$} would have been equal to approximately $17$ days}.
\end{itemize}
%
\section{Conclusion and future work} \label{sec-conc}
In this paper a heuristic is proposed for the design of uncertainty-aware dynamic output feedback for uncertain nonlinear models. This heuristic is based on off-line solution of {\bf  deterministic} NLP problems over a cloud of sampled contexts followed by a receding horizon samples collection that enriches the learning data for the same off-line computation load. The results show promising closed-loop performance when compared to ideal performances obtained with the rigorous knowledge of the parameters. These performance highly outperform nominal design based on the most expected value of the parameters. \\ \ \\ Undergoing works concern the application of the framework to more challenging examples and investigate heuristics to quickly {\em guess} the optimal triplet ($M, m, N)$ for a given uncertain nonlinear model. 
\bibliographystyle{plain}
\bibliography{stochasticMPC.bib}
\end{document}